\documentclass[final]{iopart}
\usepackage[utf8]{inputenc}
\usepackage{graphicx}
\usepackage{cite}
\usepackage[none]{hyphenat}
\usepackage{amssymb}
\usepackage{float}
\usepackage[shortcuts]{extdash}

\newcommand{\Ud}{U_\mathrm{d}}
\newcommand{\Id}{I_\mathrm{d}}
\newcommand{\Vf}{V_\mathrm{f}}
\newcommand{\td}{t_\mathrm{d}}

\usepackage{xcolor}
\newcommand{\hl}{}

\begin{document}

\title{Effect of magnetic field configuration on double layer formation and reverse discharge ignition in bipolar HiPIMS}
\author{M. Farahani$^1$, T. Kozák$^1$, A.D. Pajdarová$^1$ and J. Čapek$^1$}
\address{$^1$ Department of Physics and NTIS -- European Centre of Excellence, University of West Bohemia, Univerzitní 8, 301 00 Plzeň, Czech Republic}
\ead{adp@kfy.zcu.cz}

\begin{abstract}

The reverse discharge (RD) phenomenon in bipolar HiPIMS has been observed when a sufficiently long positive pulse is applied to the magnetron. Due to the magnetic field, electrons accumulated behind the magnetic trap are prevented from reaching the positive target. Consequently, a space charge double layer (DL) is formed between the positive target and the plasma behind the magnetic trap, leading to electron acceleration across the DL and RD ignition. This study reveals the significant impact of the magnetic field configuration on RD ignition. Experiments are performed using a Ti target involving magnetic field variation, wire probe measurements of floating potential, and optical emission spectroscopy imaging. It is found that adjusting the magnetic field to a more balanced configuration leads to earlier RD ignition, while a more unbalanced one delays or even prevents it. Specifically, the time of RD ignition decreases with an increase in the magnetic null point distance from the target. Moreover, the size and shape of optical emission in the RD varies with nearby probe placement, suggesting sensitivity to external electrodes.
    
\end{abstract}


\submitto{\PSST}

\maketitle


A bipolar HiPIMS (BP-HiPIMS) is a variant of HiPIMS discharges \cite{Gudmundsson2012, Anders2017}, where a positive voltage pulse (PP) is applied after a HiPIMS negative voltage pulse (NP). The application of the PP accelerates ions, which have remained in plasma after the NP, onto the substrate \cite{Keraudy2019}, which is beneficial for film deposition \cite{Santiago2019, Velicu2019, Tiron2019, Batkova2020b, Fernandez-Martinez2022}. An interesting phenomenon associated with BP-HiPIMS is the occurrence of a reverse discharge (RD) upon the application of a sufficiently long PP. This is identified by a sudden increase in plasma emission in front of the target\cite{Kozak2020, Law2023}, associated with a decrease in the plasma and floating potentials. A corresponding change in the energies of ions was measured by mass spectroscopy\cite{Kozak2020}. Additionally, a decrease in electron density and an increase in electron temperature have been observed during PP\cite{Law2021}.

The RD phenomenon can be attributed to the formation of a double-layer (DL) structure induced by the accumulation of electrons between the magnetic trap and the magnetic null point during PP. The magnetic field and, specifically, the magnetic mirror at the magnetron axis limits the flow of electrons to the target (now anode) while ions move freely toward the grounded walls or magnetron anode (now cathode). If the ions (e.g., Ar$^+$ ions and doubly ionized metal ions) have sufficient energy, they generate secondary electrons. As the accumulation of the electrons behind the magnetic trap continues, a drop in plasma and floating potential occurs in this region, resulting in the formation of the DL structure\cite{Tiron2020, Keraudy2019, Avino2021, Han2022, Zanaska2022}. Subsequently, electrons are accelerated across the DL's potential rise, leading to the ignition of the discharge as evidenced by the observation of ``jet-like'' or ``bulb-like'' anode light patterns (ALP) by optical emission spectroscopy (OES) imaging of the plasma \cite{Law2023, Klein2023, Pajdarova2024}. These accelerated electrons can also ionize atoms located inside the ALP core. The newly created ions together with those remaining here from the NP are accelerated in the DL's potential drop towards the substrate. It is believed that the creation of the DL structure may be beneficial for film deposition as these ions bombard the substrate with elevated energies.

In our recent paper\cite{Pajdarova2024}, we proposed that the geometry and strength of the magnetron's magnetic field play important \hl{roles} in the formation of the DL structure and the RD ignition. This can explain \hl{the} inconsistent results of plasma potential evolution obtained for balanced and unbalanced magnetic fields reported in the literature. \hl{For example, in \cite{Velicu2019, Klein2023}, the formation of the DL structure at some distance between the target and the substrate is observed as a decrease in the plasma potential, contrary to \cite{Pajdarova2020, Hippler2020}, where no such decrease is detected up to the substrate during whole PP.} The inconsistency can be attributed to the fact that each study focused on a single specific magnetic field configuration. 

In this work, a systematic investigation of the effect of the magnetic field on DL formation and RD ignition is performed using a magnetron with an adjustable magnetic field. \hl{The magnetic field topology varies between more unbalanced and more balanced configurations with respect to the standard unbalanced configuration used in \cite{Kozak2020, Pajdarova2020, Pajdarova2024}.} The onset of the RD ignition is detected by time-resolved OES imaging and by measuring the floating potential. The study confirms the connection between DL formation and RD ignition, as proposed by Pajdarová et al.\cite{Pajdarova2024}, for all studied magnetic field configurations. The timing of RD ignition is found to be strongly correlated to the magnetic field configuration, and the  \hl{positioning} of external electrodes near the discharge plasma is identified \hl{as influencing} the RD.  

\begin{figure}
    \centering
    \includegraphics[width=1\linewidth]{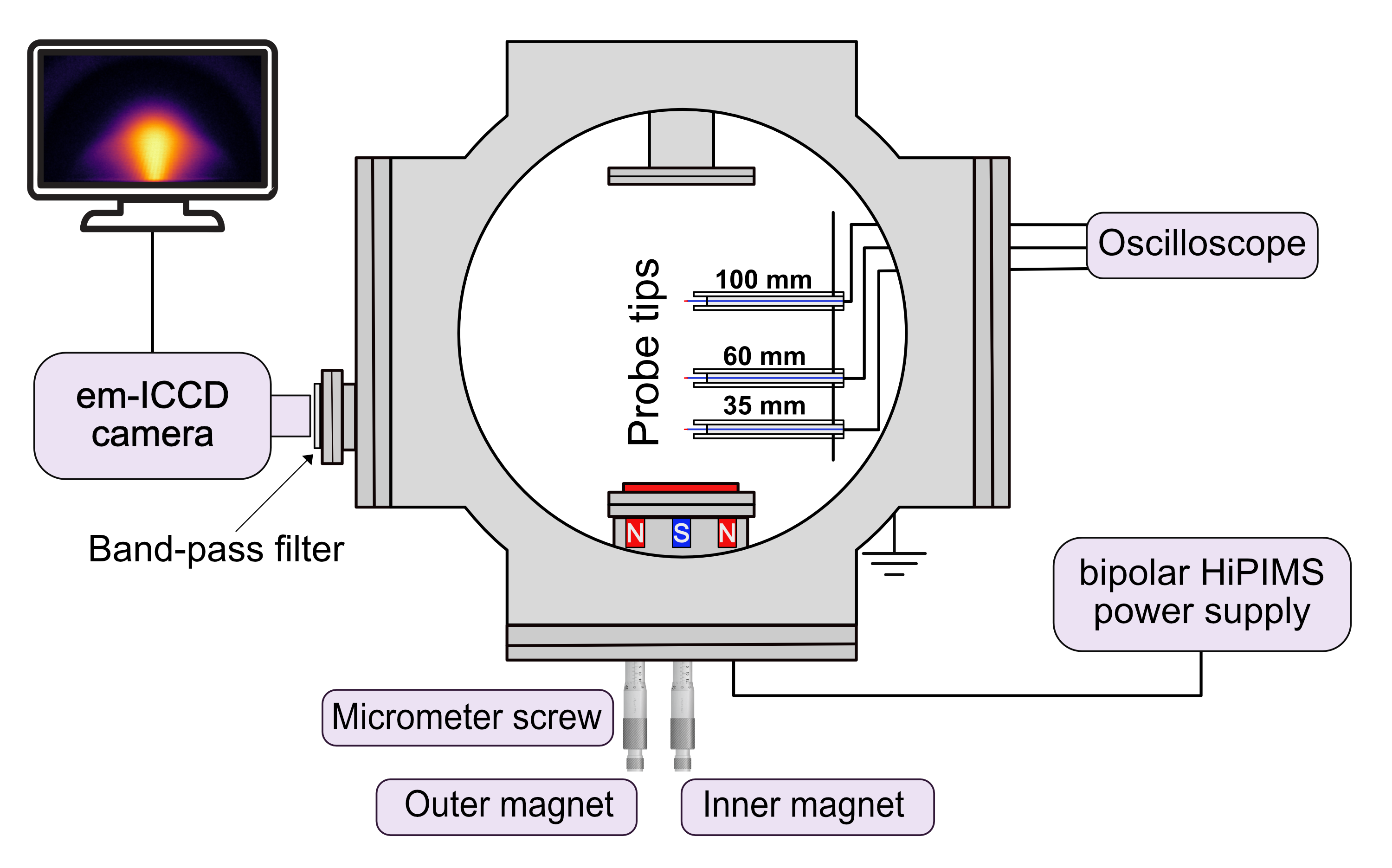}
    \caption{Schematic of the experimental setup with the emICCD camera and probes.}
    \label{fig:schema}
\end{figure}

The experimental setup (\fref{fig:schema}) employed a stainless-steel vacuum chamber built from DN200ISO-K six-way cross piping equipped with a planar circular magnetron with a variable magnetic field (VT100, Gencoa). The chamber was evacuated by a turbo-molecular pump backed up with a scroll pump down to $10^{-5}$\,Pa. The magnetron equipped with a 4" Ti target was powered by two custom-built HiPIMS pulsing units (connected to allow BP-HiPIMS operation) powered by two DC power supplies. The NP length was set to 100\,$\mu$s with a pulse-average power of 10\,kW, followed by a 5\,$\mu$s delay, while the PP pulse length was 500\,$\mu$s with a voltage amplitude of 100\,V. All experiments were carried out in pure Ar gas at a pressure of 1\,Pa \hl{and the pulse repetition frequency of 100\,Hz}.

The magnetic field of the used magnetron can be controlled by independently moving the inner (central cylindrical) and the outer (ring) magnet in the vertical direction using a micrometer screw. In this study, the magnet positions (a higher number indicates a greater distance from the target) were adjusted in two sequences when only one magnet was moved while the second one was fully inserted (i.e., at the closest position to the target). In the first sequence, the outer magnet was moved from 0 to 10\,mm with a 2\,mm step. As the magnetic field at the target edge is weakened, the configuration is changed from unbalanced to \hl{more} balanced. In the second sequence, the inner magnet was moved from 0 to 10\,mm with a 2\,mm step. This way, the magnetic field at the target center is weakened, and thus, the configuration becomes even more unbalanced. \hl{The standard unbalanced configuration used in \cite{Kozak2020, Pajdarova2020, Pajdarova2024} is with the magnets fully inserted (both at the position 0\,mm). The geometric factor $G = z_\mathrm{null} / R_\mathrm{c}$ \cite{Gencoa2001}, where $z_\mathrm{null}$ is the vertical distance of the magnetic null point from the target, and $R_\mathrm{c}$ is the target radius, characterizes the degree of magnetron unbalance and varies between 0.66 and 1.16 in our experiments.} Each measurement sequence was conducted in the shortest possible time to minimize any changes due to the target erosion. \hl{During the experiments, the target was already partially eroded (by about 1.4~mm at the deepest point of the racetrack).}

The electrical probes for plasma floating potential measurements were made of 0.1\,mm diameter Kapton\-/insulated copper wire. The bare copper was exposed only on the probe tips which were positioned parallel to the target surface and along the magnetron axis at a distance of 35, 60, and 100\,mm from the target center. The wires were led through two concentric ceramic tubes of different lengths to prevent short-circuiting by the deposited metal (\fref{fig:schema}). The target voltage and floating potentials were monitored using voltage probes (TESTEC, 100× attenuation), and the target current was monitored with a current probe (TCP303, Tektronix). All the probes were connected to two digital oscilloscopes (Picoscope 5444D, Pico Technology). The presented final waveforms are the averages from 64 consecutive periods.

\begin{figure}
    \centering
    \includegraphics[width=1\linewidth]{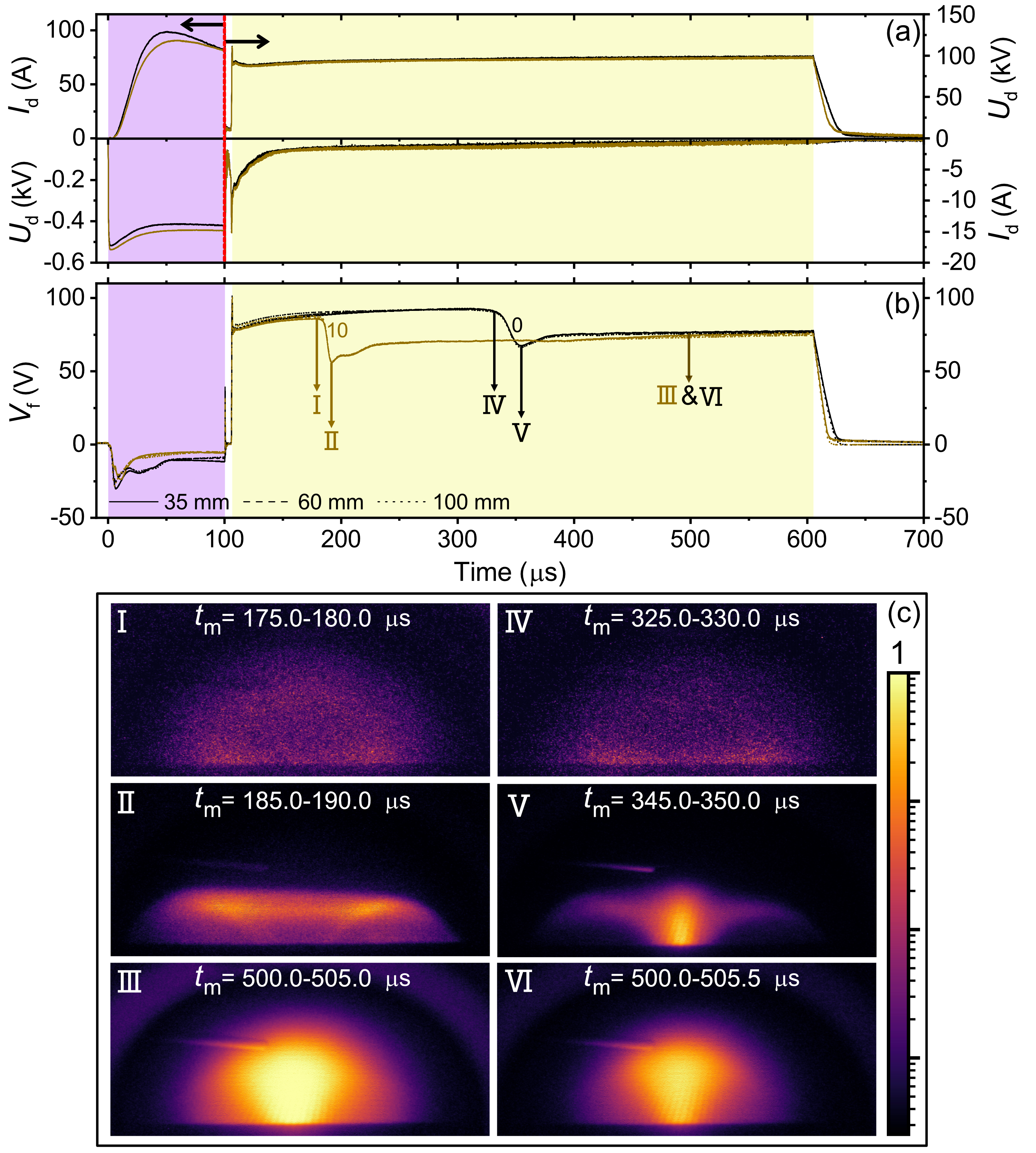}
    \caption{(a) The waveforms of the magnetron voltage, $\Ud$, and the discharge current, $\Id$, during NP (purple section, left scale) and PP (yellow section, right scale). (b) The time evolution of the floating potential, $\Vf$, at the discharge centerline for 0 and 10\,mm outer magnet positions at the distances of $z$ = 35, 60, and 100\,mm from the target. (c) The light emission patterns from Ar atoms for the selected times ($t_\mathrm{m}$ is time of measurement)  marked by Roman numbers  in (b) are shown for 10\,mm (left column, balanced magnetic field configuration) and 0\,mm (right column, unbalanced magnetic field configuration) outer magnet positions. Note that the wire probe at 35\,mm is visible in the images.}
    \label{fig:waveforms}
\end{figure}

During the PP, the light emission from Ar atoms was recorded using an emICCD camera (PI-MAX 4, SR Intensifier, Princeton Instruments) with a maximum amplification of 10000, 2 accumulation on the chip, and the gate width of 5\,$\mu$s. The final image is an average of 200 exposures. The camera, fitted with a wide-angle lens and a band-pass filter (a central wavelength of 811\,nm and a FWHM of 3\,nm), was placed perpendicular to the magnetron axis and captured images at 5\,$\mu$s intervals to cover the whole PP. The signal in captured images was normalized to be one for the maximum registered intensity, and the logarithmic scale was used to enhance the visibility of light emission patterns. To reveal the light intensity distribution at the central discharge cross-section perpendicular to the target, inverse Abel transforms of axially symmetrized images were calculated. The Gaussian blur filter (with a standard deviation of 10 pixels) was applied to the symmetrized images before the inverse Abel transform to suppress the pixel patterns of the emICCD camera (see the fringes in \fref{fig:waveforms}c).
 
\begin{figure*}
    \centering
    \includegraphics[width=1\linewidth]{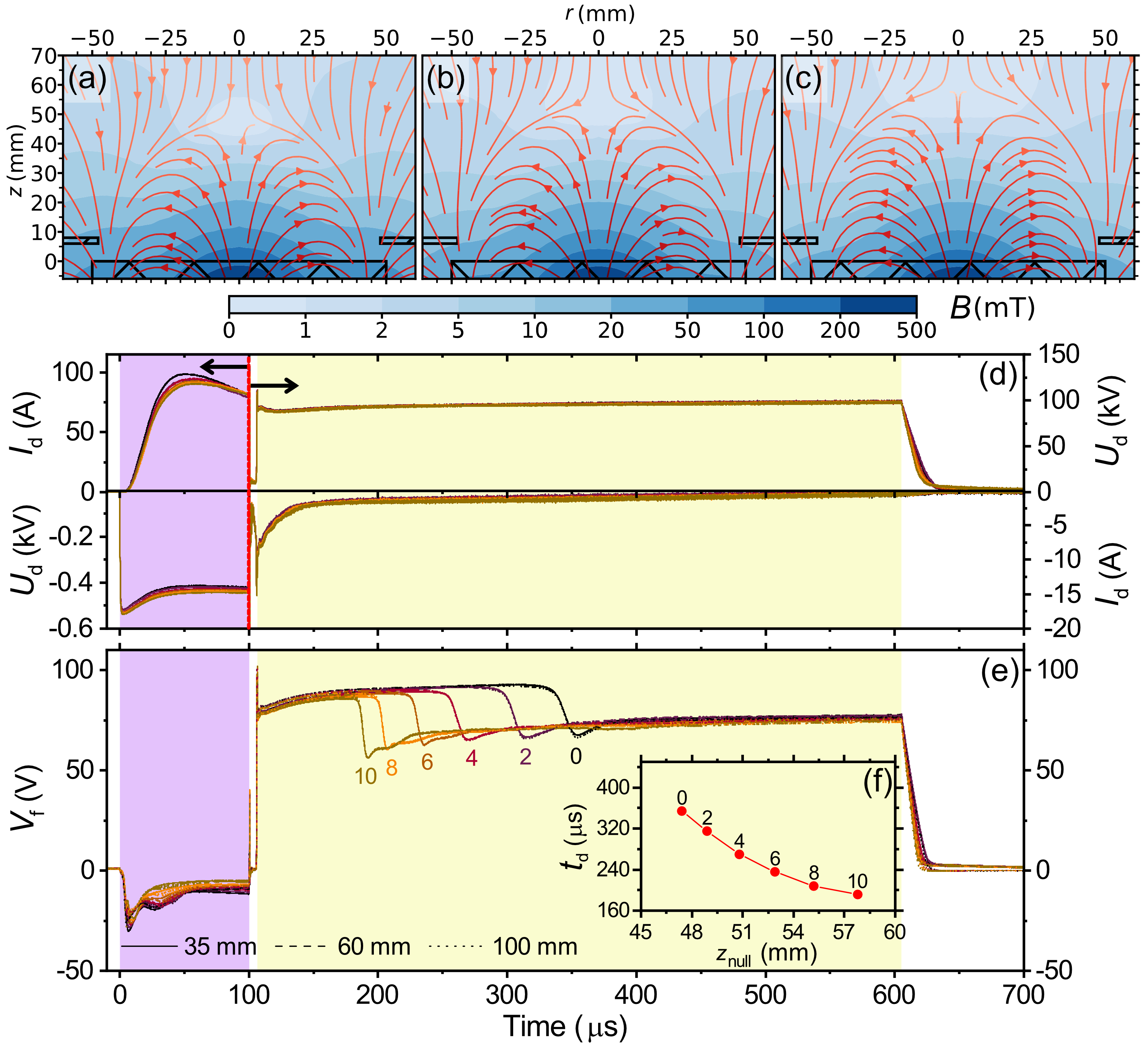}
    \caption{The magnetic field of the magnetron where $z$ is the vertical distance from the target surface, with the positions of the target and grounded anode for (a) 0\,mm, (b) 6\,mm, and (c) 10\,mm outer magnet position. The effect of the outer magnet position on (d) the waveforms of the magnetron voltage, $\Ud$, and the discharge current, $\Id$, during NP (purple section, left scale) and PP (yellow section, right scale), and (e) time evolution of the floating potential, $\Vf$, in the distances 35, 60, and 100\,mm from the target, as well as (f) the dependence of the DL creation time ($\td$) on the vertical null point distance from the target surface ($z_\mathrm{null}$).}
    \label{fig:outer}
\end{figure*}

\Fref{fig:waveforms} shows the distinction between two main magnetic field configurations: the unbalanced (outer magnet position at 0\,mm) and the \hl{more} balanced (outer magnet position at 10\,mm). Let us note that the shapes of the corresponding magnetic fields are plotted in \fref{fig:outer} (panel (a) and (c), respectively) below. Panel (a) in \fref{fig:waveforms} shows the waveforms of magnetron voltage, $\Ud$, and discharge current, $\Id$. During the NP, there is a slight difference in both $\Ud$ and $\Id$ between the unbalanced and \hl{more} balanced configurations, but they are almost identical for both configurations during the PP. Panel (b) in \fref{fig:waveforms} shows the temporal evolution of the floating potential, $\Vf$, at distances of 35, 60, and 100\,mm from the target centerline. Regarding the $\Vf$ evolution, it should be noted that there is practically no difference between the traces recorded simultaneously at the three positions for both magnetic field configurations. The most notable difference is during the NP, where $\Vf$ is slightly more negative for the unbalanced magnetic field than for the \hl{more} balanced case. 

A distinct drop of $\Vf$ occurs during the PP in both magnetic field configurations. The $\Vf$ drop accompanies a similar drop in the plasma potential (not shown here) associated with the DL formation\cite{Pajdarova2024}. This drop initiates RD ignition as confirmed by the images shown in panel (c). For the \hl{more} balanced magnetic field, the DL formation and RD ignition happen much earlier during PP. Additionally, the $\Vf$ drop for the \hl{more} balanced magnetic field is greater and faster than for the unbalanced one. After the stabilization of the $\Vf$ (around 450\,$\mu$s), its magnitude remains approximately the same for both configurations until the end of the PP.

Apart from the sooner ignition, the RD plasma emission notably differs between the two configurations, as can be seen in panel (c). Prior to RD ignition (pictures I and IV), the plasma emission was low and dispersed in the volume above the target. The images taken when $\Vf$ is at the minimum value during the PP exhibit different emission distributions: a ``fountain-like'' shape for the \hl{more} balanced configuration (II) and a ``jet-like'' shape for the unbalanced case (V). However, images taken later during the PP, at the 500\,$\mu$s (pictures III and VI), show very similar ``bulb-shaped'' emissions for both configurations. This indicates that for the \hl{more} balanced magnetic field, a more complex shape change occurs between the RD ignition until its stabilization (see the supplementary video file). Furthermore, in the case of the \hl{more} balanced magnetic field, the ALP is slightly wider.

\begin{figure*}
    \centering
    \includegraphics[width=1\linewidth]{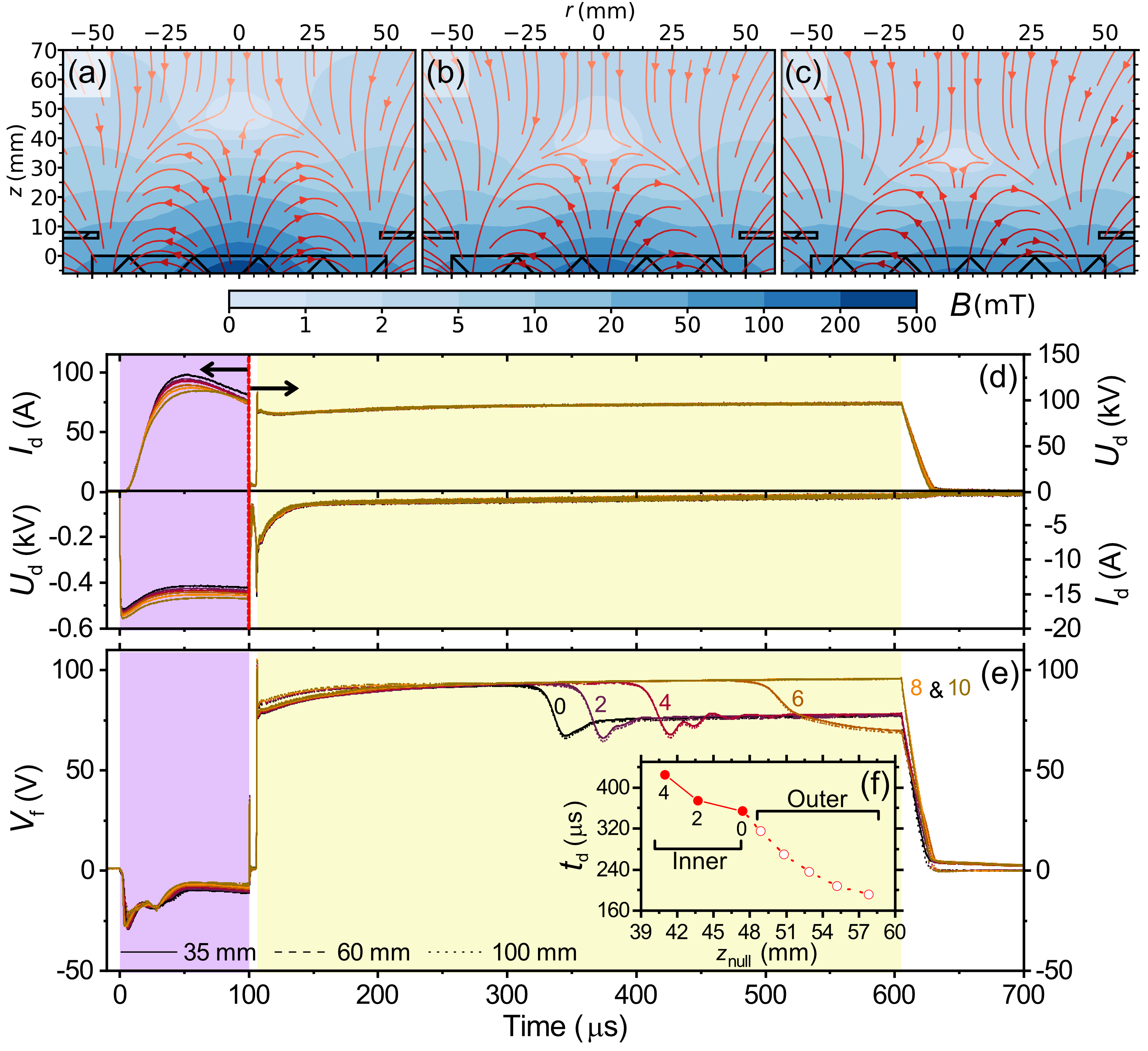}
    \caption{The magnetic field of the magnetron where $z$ is the vertical distance from the target surface, with the positions of the target and grounded anode for (a) 0\,mm, (b) 6\,mm, and (c) 10\,mm inner magnet position. The effect of the inner magnet position on the (d) $\Ud$ and $\Id$ during NP (purple section, left scale) and PP (yellow section, right scale), and (e) time evolution of the $\Vf$ in the distances 35, 60, and 100\,mm from the target, as well as (f) the dependence of the DL creation time, $\td$, on the vertical null point distance from the target surface, $z_\mathrm{null}$, \hl{(dotted line with empty circles for outer magnet position change, copied from \fref{fig:outer}f, and solid line with full circles for inner magnet position change, corresponding to $\Vf$ evolution in panel (e))}.}
    \label{fig:inner}
\end{figure*}

Figures \ref{fig:outer} and \ref{fig:inner} show the influence of the magnet position change from 0 to 10\,mm for the outer and inner magnets, respectively. The a-c panels of both figures visually represent the progressive alteration in the magnetic field geometry (for magnet positions 0, 6, and 10\,mm, in this order). It should be noted that the streamlines in the panels a-c illustrate the magnetic field geometry, not the magnetic field strength. Panels (d) show the effect of the magnetic field configuration on the waveforms of $\Ud$ and $\Id$ during the NP and PP for different magnet positions. Additionally, panels (e) present changes in the time evolution of $\Vf$ measured at three distances 35, 60, and 100\,mm from the target centerline induced by the magnetic field geometry alteration.

As shown in \fref{fig:outer}a-c, when the outer magnet is moved, the magnetic field intensity around the target centerline remains constant, but it weakens near the target edge (particularly around the anode ring), resulting in a gradual balancing of the magnetic field.

\Fref{fig:outer}d shows that the target waveforms during the NP exhibit only minor variations, and during the PP, they remain almost identical for all configurations. 

The difference in the $\Vf$, among various probe positions, is negligible, as the curves overlap for each magnetic configuration in \fref{fig:outer}e. The drop in $\Vf$, which corresponds to the ignition of the RD, is clearly shifting to lower times with increasing outer magnet position. The corresponding time of the drop event (time of ignition) is denoted as $\td$ and shown according to vertical null point distance from the target surface ($z_\mathrm{null}$) in inset (f), highlighting the monotonous decrease of $\td$ with an increasing outer magnet position. It should be noted that the time of ignition is consistent over many HiPIMS pulses and the trend of $\td$ is repeatable. Therefore, the magnetic field strongly determines the plasma conditions necessary for RD ignition. Additionally, the drop in $\Vf$ happens faster and is more pronounced for the 10\,mm position compared to the 0\,mm position. The increase in $\Vf$ after the drop exhibits roughly the same magnitude for all magnetic configurations.

As shown in \ref{fig:inner}a-c, an increase in the position of the inner magnet results in a weaker magnetic field around the target center while the strength of the magnetic field around the target edge remains unchanged. Consequently, the magnetic field changes from the unbalanced to a more unbalanced configuration, and the $z_\mathrm{null}$ decreases. Just as for the variation of the outer magnet position, $\Ud$ and $\Id$ waveforms in \fref{fig:inner}d exhibit the same trends. The drop in $\Vf$, see \fref{fig:inner}d, appears later as the inner magnet position increases from 0 to 6\,mm. For the positions of 8 and 10\,mm, the drop event does not occur during the PP. As the drop shifts to later times, $\Vf$ also decreases slower during the drop event. For the 6\,mm position, a gradual decrease in $\Vf$ is observed without a subsequent increase, as the PP terminates shortly after the drop. In the other cases, the increase in $\Vf$ after the drop is approximately of the same magnitude for all magnet positions where the RD occurred. In the inset (f), the monotonous increase of $\td$ with increasing inner magnet position is emphasized. \hl{The dotted line with the empty circles shows the change of $\td$ with the outer magnet position (copied from figure \ref{fig:outer}f), while the solid line with full circles shows the variation with the inner magnet position (as indicated in \ref{fig:inner}e). The inset thus shows the continuous increase in $\td$ as the magnetic configuration becomes more unbalanced ($z_\mathrm{null}$ decreases), first by moving the outer magnet from 10~mm to 0~mm, and then by moving the inner magnet from 0~mm to 10~mm.}

Our explanation of the physical mechanisms behind the observed phenomena is as follows. After applying the positive potential on the target, ions within the magnetic trap, that were not lost during the delay between NP and PP, flow mainly to the grounded magnetron anode (now cathode) due to its close proximity. The ions behind the magnetic trap form a diffuse bulk plasma and flow towards all nearby ground surfaces (magnetron anode, substrate, and chamber walls)\cite{Han2022, Pajdarova2024}. Upon impact, any ion with sufficient energy induces secondary electron emission from the grounded surfaces.

The electrons in the bulk plasma can easily flow only along the magnetic field lines, but the field lines converging towards the center and edge magnets form a magnetic mirror \cite{Dinklage2005a} which can also partly reduce the flux of electrons to the target, especially at the target center where the magnetic field is strongest. Only the electron that has the angle, $\Phi$, between its velocity and the local magnetic field line lower than 
\begin{equation}\label{e:LossCone}
    \Phi_\mathrm{c} = \arcsin \left ( B_0 / B_\mathrm{s} \right )^{1/2}
\end{equation}
(its velocity is in the so-called loss cone, $\Phi < \Phi_\mathrm{c}$) at the position of its last collision event (a radical change of its velocity direction) may be absorbed by the target (neglecting the acceleration by the electric field). In equation \eref{e:LossCone}, $B_0$ and $B_\mathrm{s}$ are the magnetic field strengths at the last collision event and at the point of absorption on the target, respectively. Note that the loss cone does not change much during presented changes of the magnetic field ($\Phi_\mathrm{c}$ is between 36.4$^\circ$ and 38.6$^\circ$ at the distance of 1\,cm above the target for an electron following the magnetic field line in the magnetron center).

Since the plasma potential is practically equal to the target voltage in the first part of the positive pulse (before the drop event), there must be a path for the electrons to reach the target and balance the flux of ions to grounded surfaces and the emission of secondary electrons. Most electrons likely flow towards the target edge where a path along the magnetic field lines is open\cite{Pajdarova2024}, especially for the unbalanced configuration. The simulations of Han et. al \cite{Han2022} indicate that the plasma density drops faster near the grounded target anode due to the density and electric potential gradients. Consequently, the plasma potential decreases around the anode as its sheath extends until the path for the electrons toward the target surface is blocked. Then, the plasma potential decreases in the whole volume outside the magnetic trap, which is observed here by the drop in the floating potential, and the DL is formed. Acceleration of the electrons within the DL leads to ionization and excitation of species, causing a considerable increase in light emission (Figure \ref{fig:waveforms}c \hl{II-III and V-VI}), which indicates the RD ignition and coincides with the $\Vf$ drop.

As mentioned earlier, increasing the outer magnet position weakens the magnetic field near the target edge (Figure \ref{fig:outer}a-c) and changes the magnetic configuration from unbalanced to \hl{more} balanced. As shown (Figure \ref{fig:outer}c), the path for electrons to the target is narrower since the grounded anode ring repels electrons. This leads to earlier ignition of the RD as the accumulation of electrons behind the magnetic trap is faster. Conversely, in the unbalanced magnetic configuration, the magnetic field lines allow a more unrestricted flow of electrons to the target edge. 

Increasing the inner magnet position weakens the magnetic field at the target center and changes the configuration to even more unbalanced. However, evaluation of \eref{e:LossCone} reveals that the loss cone angle is not significantly reduced. So, the dominant effect responsible for the observed delaying of the RD ignition is likely again the shape of the magnetic field determining the flow of electrons towards the target edge. For the inner magnet positions of 8 and 10\,mm, as the field is very unbalanced and the center magnetic field is weak, the RD ignition is not triggered during the 500\,$\mu$s long PP.

The inset in \fref{fig:inner}f shows a monotonous dependence of the RD ignition time and the magnetic field null point position. This indicates that the degree of unbalance of the magnetic field is the dominant factor determining the RD ignition. On the other hand, a change in the trend is visible at the point joining the two series of experiments (inner and outer magnet position) indicating some role of the absolute strength of the magnetic field at the target center.

\begin{figure}
    \centering
    \includegraphics[width=1\linewidth]{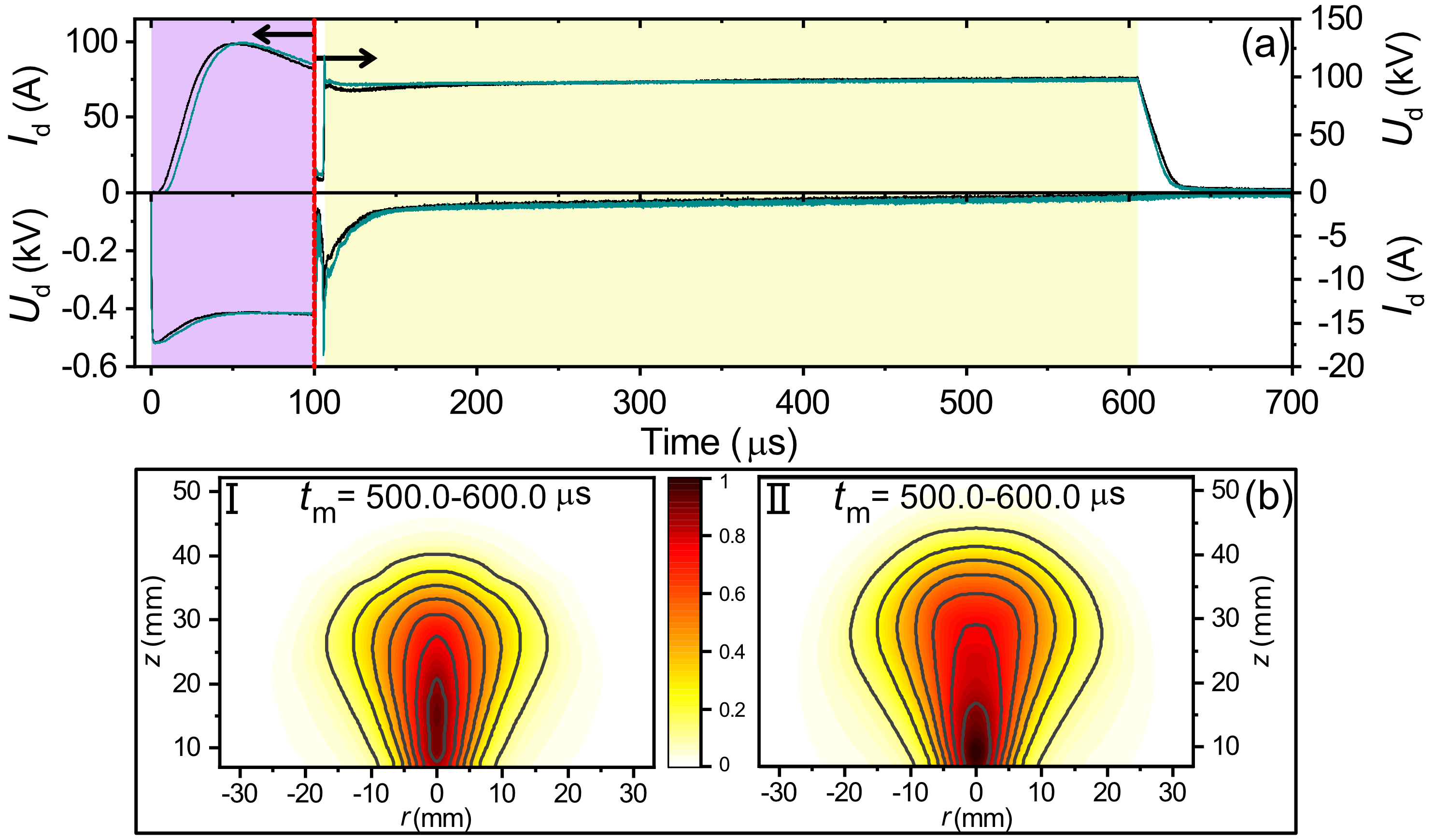}
    \caption{(a) The waveforms of the $\Ud$ and $\Id$ during NP (purple section-left scale) and PP (yellow section-right scale) for both with probes (black line) and without probes (green line). (b) The light intensity distribution on the central cross-section perpendicular to the target calculated by the inverse Abel transforms of recorded light emission averaged for the time interval from 500\,$\mu$s to 600\,$\mu$s with (I) and without probes (II).}
    \label{fig:probe}
\end{figure}

Figure \ref{fig:probe} shows that despite the target current and voltage waveforms are practically identical (see panel (a)), the distribution of light emission from Ar atoms recorded with (I) and without the three wire probes (II) (panel (b)) is different. Specifically, in the presence of the probes, the plasma emission profile is narrower and more compressed towards the target (see also the supplementary video file). This confirms that the RD evolution can be slightly affected by the presence of measurement probes or, in general, by other surfaces placed in proximity to the plasma. Specifically, the secondary electron emission from the probe stems covered by the sputtered metal or other surfaces can affect the RD ignition. A possible influence of plasma by the metal-covered probe stems and their positioning systems may also explain the differences in the direction of the apparent movement of the plasma and floating potential drops observed in the literature during the DL formation (e.g., from the target to the substrate \cite{Zanaska2022, Pajdarova2024}, and in the opposite direction \cite{Avino2021, Law2021}).

To conclude, the investigation of the bipolar HiPIMS discharge provided crucial insights into the relationship between magnetic field configuration and the ignition of the reverse discharge. A direct correlation between the drop in floating (and plasma) potential, indicative of double layer formation, and RD ignition, as confirmed by an increase in optical emission in the form of an anode light pattern, was revealed. The time of RD ignition exhibits a monotonous dependence on the magnetic null point distance. A more balanced magnetic field leads to earlier ignition, while unbalanced configurations lead to delayed ignition or even suppression. Notably, the floating potential behind the magnetic trap remained practically identical across different probe positions, indicating a simultaneous potential drop at different distances from the target. Additionally, the size and shape of anode light patterns were observed to vary with nearby probe placement, suggesting sensitivity to external electrodes.

\section*{Acknowledgments}

This work was supported by the project QM4ST with No. CZ.02.01.01/00/22\_008/0004572, co-funded by the ERDF as part of the Ministry of Education, Youth and Sport.

\section*{References}

\bibliographystyle{iopart-num}
\bibliography{references}

\end{document}